\documentclass[%
 aps,
twocolumn,
superscriptaddress,
pra,
amsmath,amssymb
]{revtex4-1}

\usepackage{graphicx}% Include figure files
\usepackage{dcolumn}% Align table columns on decimal point
\usepackage{bm}% bold math
\usepackage{color}
\usepackage{hyperref}
\begin{document}

\preprint{APS/123-QED}

\title{Detecting Casimir torque with an optically levitated nanorod}% Force line breaks with \\
%\thanks{A footnote to the article title}%
\author{Zhujing Xu}
\affiliation{Department of Physics and Astronomy, Purdue University, West Lafayette, Indiana 47907, USA}
\author{Tongcang Li}
\email{Corresponding author: tcli@purdue.edu}
\affiliation{Department of Physics and Astronomy, Purdue University, West Lafayette, Indiana 47907, USA}
\affiliation{School of Electrical and Computer Engineering, Purdue University, West Lafayette, Indiana 47907, USA}
\affiliation{Purdue Quantum Center, Purdue University, West Lafayette, Indiana 47907, USA}
\affiliation{Birck Nanotechnology Center, Purdue University, West Lafayette, Indiana 47907, USA}

\date{\today}% It is always \today, today,
             %  but any date may be explicitly specified

\begin{abstract}
The linear momentum and angular momentum of virtual photons of quantum vacuum fluctuations can induce the Casimir force and the Casimir torque, respectively. While the Casimir force has been measured extensively, the Casimir torque has not been observed experimentally though it was predicted over forty years ago. Here we propose to detect the Casimir torque with an optically levitated nanorod near a birefringent plate in vacuum. The axis of the nanorod tends to align with the polarization direction of the linearly polarized optical tweezer. When its axis is not parallel or perpendicular to the optical axis of the birefringent crystal, it will experience a Casimir torque that shifts its orientation slightly.
We calculate the Casimir torque and Casimir force acting on a levitated nanorod near a birefringent crystal.  We also investigate the effects of thermal noise and photon recoils on the torque and force detection. We prove that a levitated nanorod in vacuum will be capable of detecting the Casimir torque under realistic conditions, and will be an important tool in precision measurements.

%\begin{description}
%\item[PACS numbers]
%May be entered using the \verb+\pacs{#1}+ command.
%\end{description}
\end{abstract}

\pacs{Valid PACS appear here}% PACS, the Physics and Astronomy
                             % Classification Scheme.
%\keywords{Suggested keywords}%Use showkeys class option if keyword
                              %display desired
\maketitle

%\tableofcontents
\section{Introduction}
%\begin{figure}[b,p]
\begin{figure}
  \centering
    \includegraphics[width=0.5\textwidth]{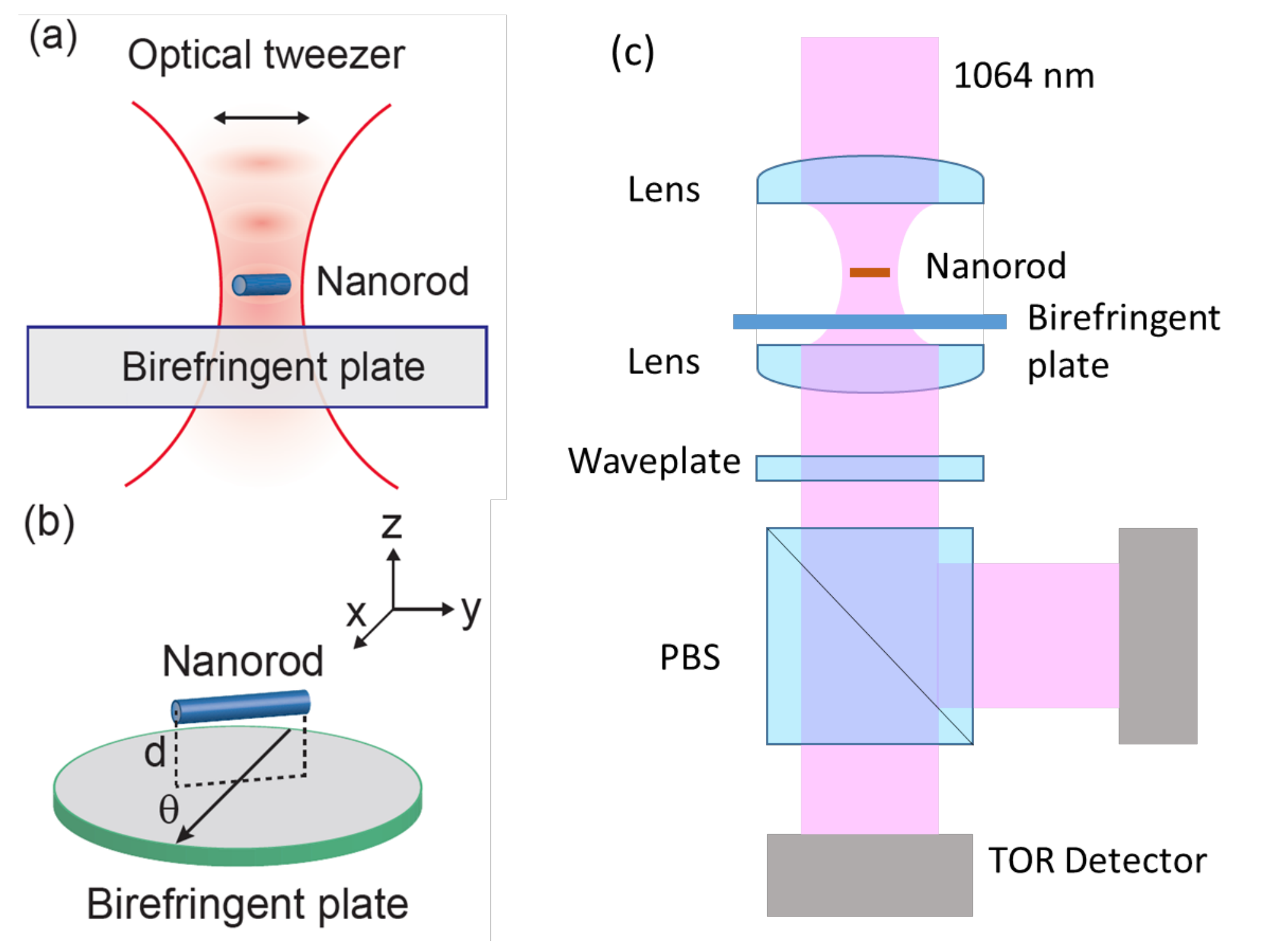}
      \caption{\label{setup}(a) A nanorod levitated by a linearly polarized optical tweezer in vacuum near a birefringent plate. Its axis tends to align with the polarization direction of the optical tweezer. (b) There will be a Casimir torque on the nanorod when its axis is not aligned with a principle axis (the black arrow) of the birefringent plate. The trapping laser beam can also be used as the detecting beam to measure the torsional motion of the nanorod. Here $d$ is the separation between the nanorod and the plate, and $\theta$ is the angle between the long axis of the nanorod and the optical axis of the birefringent plate. (c) A proposed experimental scheme for detecting torsional (TOR) vibration of a levitated nanorod. A silica nanorod (represented by a brown rod) is levitated by a tightly focused linearly polarized 1064 nm laser beam and it is very close to a birefringent plate. The angle of the nanorod is monitored by the exiting trapping laser. A tunable waveplate balances the power of the beams after the polarizing beam splitter (PBS) for the TOR detector.}
\end{figure}

A remarkable prediction of quantum electrodynamics (QED) is that there are an infinite number of virtual photons in vacuum due to the zero-point energy that never vanishes, even in the absence of electromagnetic sources and at a temperature of absolute zero. In 1948, Casimir predicted an atrractive force between two ideal metal plates due to the linear momentum of virtual photons \cite{Casimir}. The number of electromagnetic modes between two metal plates is less than the number of modes outside the plates, thus the plates experience an attractive force, which is Casimir force. Casimir force has already been measured many times throughout the years \cite{Casimir force measurement, Casimir force 1998, Casimir Decca 2003, repulsive Casimir force, Yukawa Decca 2016, Jacob book, casimir review, Munday, V.Mostepanenko}. Besides the linear momentum, the angular momentum carried by virtual photons can generate the Casimir torque (or van der Waals torque) for anisotropic materials\cite{Parsegian, Barash, Enk1995}.
Despite significant interests about the van der Waals and Casimir torque \cite{Siber, Siber2, PSSB201147150, Munday2, Munday3, Rajter, Munday5, Yasui2015, Zhang2017science}, the torque has not been measured experimentally though it was predicted over 40 years ago, mainly due to the lack of a suitable tool \cite{casimir review, Munday, V.Mostepanenko, Jacob book}.

Here we propose a method to measure the Casimir torque with a nanorod levitated by a linearly polarized optical tweezer in vacuum near a birefringent plate. The relative orientation between the nanorod and the birefringent crystal could be manipulated by the polarization of the trapping laser beam. When the long axis of the nanorod is not aligned with a principle axis of the birefringent plate, there will be a Casimir torque acting on the nanorod, which tries to minimize the energy, as shown in FIG.~\ref{setup}.(a),(b). Here $d$ is the separation between the nanorod and the plate, and $\theta$ is the angle between the long axis of the nanorod and the optical axis of the birefringent plate. Casimir torque is related to separation $d$ and relative orientation $\theta$, which will be shown in the next section.

An optically levitated nanoparticle in vacuum can have an ultrahigh mechanical quality factor ($Q>10^{9}$) as it is well-isolated from the thermal environment, which is excellent for precision measurements \cite{Yin2014, Li, Li2010, Levitated nanorod, Cavity mechanics, Levitated nanorod2, Levitated nanorod3, Full rotational, Ro-translational, Geraci2010, feedbackcooling, forcesql, zeptonewtonforce}. Optical levitation of a silica (SiO$_2$) nanosphere in vacuum at $10^{-8}$ torr \cite{forcesql}, and  force sensing at  $10^{-21}$N level with a levitated nanosphere \cite{zeptonewtonforce} have been demonstrated in two separate experiments.  The libration of an optically levitated nonspherical nanoparticle in vacuum has also been observed \cite{Li, Full rotational}, which provides a solid foundation for this proposal. 
We are going to detect the torsional vibration of the nanorod and measure its orientation with the laser polarization in a scheme similar to those reported in Ref. \cite{Li, Full rotational}.
The nanorod will be levitated using an optical tweezer formed by a linearly polarized 1064 nm laser beam near a birefringent plate (FIG.~\ref{setup}).  The torsional vibration of the nanorod will dynamically change the polarization of the laser
beam, which can be detected with a polarizing beam splitter (PBS) and a balanced detector. The birefringent plate will cause a static change of the polarization of the laser, which can be canceled by a tunable waveplate as shown in FIG.~\ref{setup}.(c).

In this paper, we will show that a silica nanorod with a length of $200$~nm and a diameter of $40$~nm levitated by a 100 mW optical tweezer in vacuum at $10^{-7}$ torr will have torque detection sensitivity about $10^{-28}~{\rm Nm}/\sqrt{\rm Hz}$ at room temperature. The Casimir torque between a nanorod with the same size and a birefringent plate separated by 266 nm is calculated to be on the order of $10^{-25}~{\rm Nm}$. The Casimir torque is 3 orders of magnitude larger than the minimum torque we can detect in 1 second, and thus will be detectable with our system.
A levitated nanorod in vacuum will be several orders more sensitive than the state-of-the-art torque sensor. The best reported torque sensitivity is $2.9\times 10^{-24}~{\rm Nm}/\sqrt{\rm Hz}$, which was achieved by cooling a cavity-optomechanical torque sensor to 25mK  in a dilution refrigerator \cite{torquesql}. The force detection sensitivity will be limited by the thermal noise when the pressure is above $10^{-7}$ torr. When the pressure is below $10^{-7}$ torr, the force sensitivity is mainly limited by photon recoil, which is about $10^{-21}~\rm N/\sqrt{\rm Hz}$. Our calculated turning point of the force sensitivity around $10^{-7}$ torr is consistent with the experimental observation of photon recoils around $10^{-7}$ torr \cite{forcesql}. The exact turning point depends on the size and shape of the nanoparticle, as well as the intensity of the trapping laser.

Compared to the recent proposal of detecting the effects of Casimir torque with a liquid crystal \cite{Munday5}, our method with a levitated nanorod in vacuum will be able to measure the Casimir torque at a much larger separation ($d>200~\rm nm$), where retardation is significant. We will also be able to investigate the Casimir torque as a function of relative orientation in detail.  As an ultrasensitive nanoscale torsion balance \cite{torquesql},  our system 
will also enable many other precision measurements.

\section{Trapping potential}
We consider a silica  nanorod with a length of $l=200$ nm in the long axis and a diameter of  $2a=40$ nm trapped with a linearly polarized Gaussian beam in vacuum.
The electric field of the beam can be described under the paraxial approximation as
\begin{eqnarray}\label{eq:electricfield}
E_{x}(x,y,z)=E_{0}\frac{\omega_0}{\omega(z)}
exp\{\frac{-(x^2+y^2)}{[\omega(z)]^2}\}\nonumber\\
\times exp(ikz+ik\frac{x^2+y^2}{2R(z)}-i\zeta(z))\hspace{0.1cm},\\
E_{y}(x,y,z)=E_{z}(x,y,z)=0\hspace{0.1cm} ,
\end{eqnarray}
where $E_0$ is the electric field amplitude at the origin, $\omega(z)$ is the radius at which the field amplitudes fall to $1/e$ of their axial values at the plane z along the beam, $R(z)$ is the radius of curvature of the beam's wavefronts at z and $\zeta(z)$ is the Gouy phase at z.
\begin{figure}[b]
\includegraphics[width=0.6\textwidth]{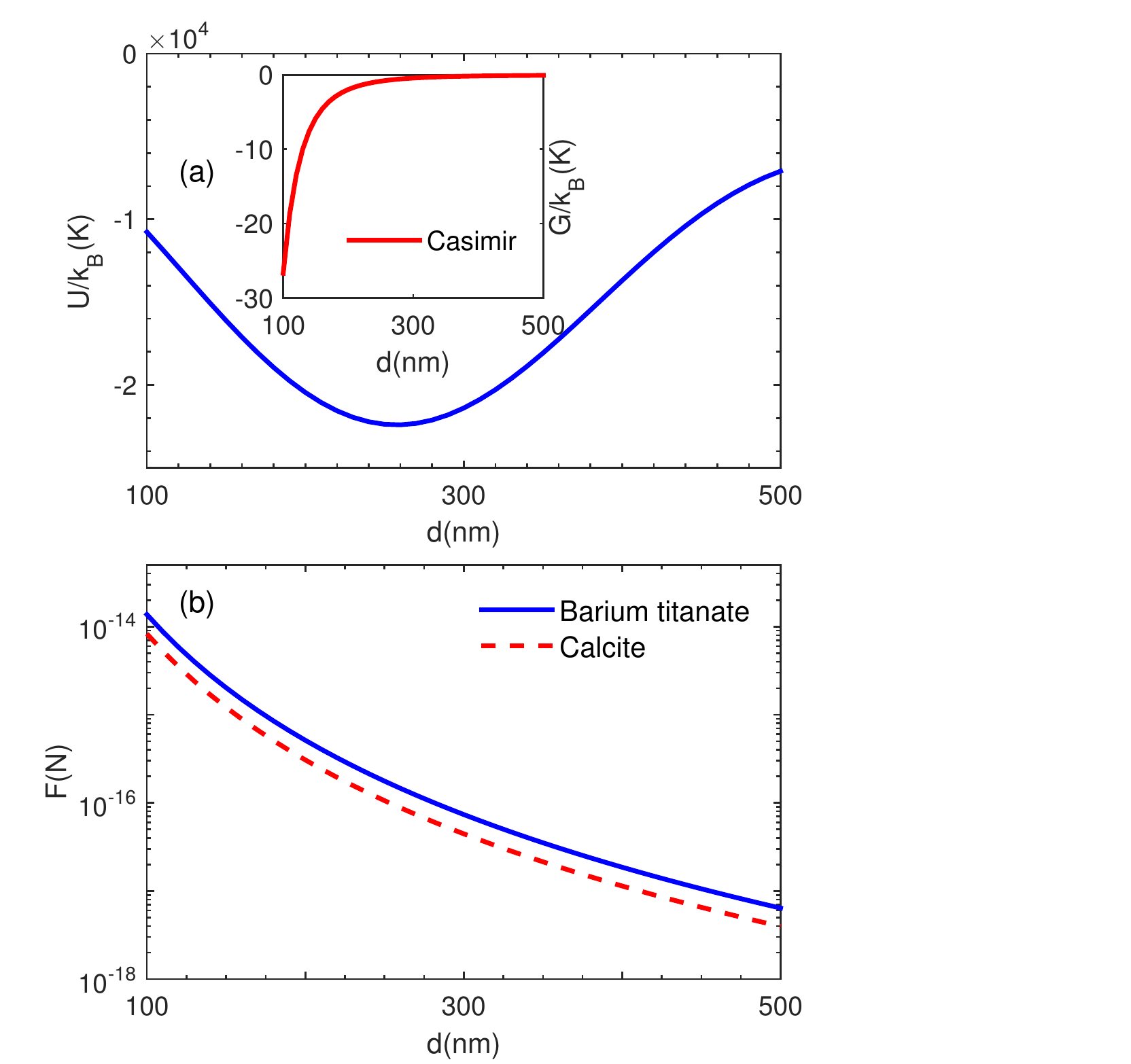}% Here is how to import EPS art
\caption{\label{potentialandforce} (a) Calculated trapping potential for a silica nanorod with a length of $200~\rm nm$ and a diameter of $40~\rm nm$ as a function of separation $d$. Here we set the center of Gaussian beam at $d=266~\rm nm$, laser power as 100~mW and laser waist radius as 400~nm. Inset: Calculated Casimir free energy at 300~K as a function of separation $d$ at relative orientation $\theta=\pi/4$. Here we choose the substrate material to be barium titanate. (b) Calculated Casimir force at 300~K as a function of separation $d$ between the silica nanorod and a birefringent crystal at relative orientation $\theta=\pi/4$. The blue solid line and red dashed line represent the Casimir force when the birefringent crystals are barium titanate and calcite, respectively.}
\end{figure}
In the case of rods with a large apsect ratio, the components of the polarizability tensor \cite{Ro-translational} parallel and perpendicular to the symmetry axis are $\alpha_{\parallel}=V\epsilon_{0}(\epsilon_r-1)$ and $\alpha_{\perp}=2V\epsilon_0(\epsilon_r-1)/(\epsilon_r+1)$. Here V is the volume of the object and $\epsilon_{r}$ is the relative permittivity of the object. The absorption of electromagnetic field of the silica is negligible, so we assume $\epsilon_r$ is real.

When the size of the nanorod is much smaller than the wavelength of the laser (here we choose wavelength as 1064 nm), we can apply the Rayleigh approximation. The induced dipole will be $\mathbf{p}=\alpha_xE_x\mathbf{\hat{x}}_N+\alpha_yE_y\mathbf{\hat{y}}_N+\alpha_zE_z\mathbf{\hat{z}}_N$, where the instantaneous electric field of the laser beam $\mathbf{E}$ is decomposed into components along the principle axes of the nanorod. The long axis of the nanorod will tend to align with the polarization of the laser.
When the vibration amplitude is small, the optical potential is harmonic around the laser focus and the vibrations of the trapped nanorod along different directions are uncoupled.
Here we focus on its   center-of-mass motion along z axis and its rotation around z axis.
Thus the potential energy of the nanorod in the optical tweezer is
\begin{eqnarray}\label{eq:potential0}
U(z,\phi)=-\frac{1}{2c\epsilon_0}[\alpha_{\parallel}-(\alpha_{\parallel}-\alpha_{\perp})\sin^2\phi)]I_{laser}(z)\hspace{0.1cm},
\end{eqnarray}
where $c$ is the speed of light, $\epsilon_0$ is the vacuum permittivity, $\phi$ is the angle between the long axis of the nanorod and the polarization of the Gaussian beam, and $I_{laser}(z)$ is the laser intensity at the location of the nanorod. In free space, the peak laser intensity at the focus is given by $I_{laser}=Pk_0^2{\rm NA}^2/(2\pi)$, where $P$ is the laser power, NA is the numerical aperture of the objective lens, and $k_0$ is the magnitude of the wave vector of the laser beam. We assume NA$=0.85$ in this paper.

Here we also need to consider the reflection from the substrate, which will form a standing wave with the incident wave and strengthen the trapping potential at the 1/4 wavelength point (FIG.~\ref{setup}). We assume that the center of Gaussian beam is 1/4 wavelength away from the surface of the substrate. In our case, the wavelength is 1064~nm, so the center of the beam will be 266~nm from the substrate surface. The refractive index is $n_0=1$ for vacuum and are $n_o=2.269$ and $n_e=2.305$ for ordinary and extraordinary axis of the birefringent crystal $\rm BaTiO_3$ at 1064~nm\cite{barium index}, respectively. If the laser is perpendicular to the surface, the reflectance along ordinary and extraordinary axis are $R_o=0.16$ and $R_e=0.15$. For a laser beam focused by a NA$=0.85$ objective lens, the angular aperture is $58^\circ$. The angle of incidence varies a lot at the surface of birefringent crystal
 and then the reflectance will become location and orientation dependent. However, only reflected wave from light with a small  incident angle will interfere with the incident wave and contributes to the trapping potential near $z$ axis, where the nanorod is trapped. Furthermore, when a  linearly polarized laser beam is focused before hitting a surface, 50\% of the laser will be parallel to the incident plane ($p$-polarized), while the other 50\% of the laser will be perpendicular to the incident plane ($s$-polarized). The average reflectance of a 50\%  $p$-polarized and 50\%  $s$-polarized laser at the maximum incident angle $58^\circ$ is 0.19, which is still close to  $R_o=0.16$ and $R_e=0.15$.
 Therefore, we use $R_o$ and $R_e$ in the calculation for simplicity.

We assume that the linearly polarized laser is polarized at $\pi/4$ relative to both optical axes of the birefringent crystal, and the axis of the nanorod is aligned with the polarization of the Gaussian beam ($\phi \approx 0$). Then the potential near $z=0$ (or $d=d_0$) is 
\begin{eqnarray}\label{eq:potential1}
U(z)=U(d-d_0)\approx -\frac{1}{4}\alpha_{\parallel}E_0^2 [\frac{\omega_0^2}{[\omega(d-d_0)]^2}\nonumber\\
+\frac{R_o+R_e}{2}
\frac{\omega_0^2}{[\omega(d+d_0)]^2}
+(\sqrt{R_o}+\sqrt{R_e}) \cos(2kd)\nonumber\\
\frac{\omega_0^2}{\omega(d-d_0)\omega(d+d_0)}]\hspace{0.1cm},
\end{eqnarray}

where $d_0=266~\rm nm$ is the distance from the center of the Gaussian beam to the birefringent crystal.
We use Eq.~\ref{eq:potential1} to calculate the trapping potential and the result is shown in FIG.~\ref{potentialandforce} (a). Here the laser power is 100~mW and the waist radius is approximately 400~nm.  The potential energy at the center of the beam is around $-2.2\times 10^4 $K, which allow us to avoid losing the nanorod from thermal motion at room temperature.

%\vspace{0.3cm}

\section{Casimir Interaction}

The Casimir force and the van der Waals force have the same physical origin, as they both arise from quantum fluctuations. Casimir forces between macroscopic surfaces involve separations typically larger than $100~\rm nm$ where retardation effect plays an important role, while van der Waals forces often refer to separations smaller than a few nm where retardation is negligible\cite{casimir review, Munday, V.Mostepanenko, Jacob book}. To calcualte the Casimir interaction between a nanorod and a birefringent plate, we follow the method in Ref. \cite{Siber, vdW book} by assuming that a half space is a dilute assembly of anisotropic cylinders. With that we could extract the interaction between a cylinder and one semi-infinite half space from the interaction free energy between two half spaces.  We notice that Ref. \cite{Siber} has two typos. In Eq. 5 about the function $N$ in Ref. \cite{Siber}, the first term in the third line, should be $\boldsymbol{\rho_3^2}(\epsilon_3-\epsilon_{1,\perp})(Q^2+\rho_{1,\perp}\rho_3)$ instead of $\boldsymbol{\rho_3^3}(\epsilon_3-\epsilon_{1,\perp})(Q^2+\rho_{1,\perp}\rho_3)$. In Eq.7  for the function $\tilde f(\phi)$ in Ref. \cite{Siber}, the term inside the square root should be $Q^2((\epsilon_{1,\parallel}/\epsilon_{1,\perp})-1)\cos^2\phi+\boldsymbol{\rho_{1,\perp}^2}$ instead of $Q^2((\epsilon_{1,\parallel}/\epsilon_{1,\perp})-1)\cos^2\phi+\boldsymbol{\rho_{1,\parallel}^2}$. The corrected interaction free energy per unit length of the cylinder, $g(d,\theta)$, between a single cylinder and a half-space substrate is

\begin{eqnarray}\label{eq:freeenergy}
g(d,\theta)=\frac{k_BTa^2}{4\pi}\sum\limits_{n=0}^\infty{'}\int_{0}^{\infty}QdQ\int_{0}^{2\pi}d\phi\left[e^{-2d\rho_3}\frac{N}{D}\right]\hspace{0.1cm},\qquad
\end{eqnarray}
where
\begin{widetext}
\begin{align}\label{eq:bigN}
N &=(\frac{\Delta_{\parallel}}{2}-\Delta_{\perp})\{Q^2 \sin^2(\phi+\theta)\times[\tilde{f}(\phi)\epsilon_{1,\perp}(Q^2 \sin^2\phi(\rho_{1,\perp}+\rho_3)+\rho_{1,\perp}\rho_3(\rho_3-\rho_{1,\perp}))
+(\epsilon_{1,\perp}-\epsilon_3)(\rho_3(\rho_{1,\perp}+2\rho_3)\nonumber\\
&-Q^2)]-2\tilde{f}(\phi)\epsilon_{1,\perp}\rho_{1,\perp}\rho_{3}^2[2Q^2 \sin{\phi} \cos{\theta} \sin(\phi+\theta)+\rho_{3}^2 \sin^2{\theta}]+
\tilde{f}(\phi)\epsilon_{1,\perp}\rho_{3}^2[Q^2\sin^2{\phi}(\rho_{1,\perp}-\rho_3)
+\rho_{1,\perp}\rho_{3}(\rho_{1,\perp}+\rho_{3})]\nonumber\\
& +\rho_{3}^2(\epsilon_{3}-\epsilon_{1,\perp})(Q^2+\rho_{1,\perp}\rho_3)\}+
2\tilde{f}(\phi)\Delta_{\perp}\epsilon_{1,\perp}[Q^2\sin^2{\phi}(Q^2\rho_{1,\perp}-\rho_3^3)+\rho_{1,\perp}\rho_{3}^2(Q^2\cos(2\phi)+\rho_{1,\perp}\rho_3)]\nonumber\\
& -\Delta_{\perp}(\epsilon_{1,\perp}-\epsilon_3)\times[(Q^2+\rho_3^2)(Q^2+\rho_{1,\perp}\rho_{3})+(Q^2-\rho_3^2)(Q^2-\rho_{1,\perp}\rho_3)]
\end{align}
\end{widetext}
and
\begin{eqnarray}\label{eq:parameterD}
D=\rho_3(\rho_{1,\perp}+\rho_3)\{\epsilon_{1,\perp}\tilde f(\phi)[Q^2\sin^2\phi-\rho_{1,\perp}\rho_{3}]\nonumber\\
+\epsilon_{1,\perp}\rho_3+\epsilon_3\rho_{1,\perp}\}\hspace{0.1cm}.\quad
\end{eqnarray}
In the equations above,
\begin{eqnarray}\label{eq:parameter1}
\rho_{1,\perp}=\sqrt{Q^2+\frac{\epsilon_{1,\perp}\omega_n^2}{c^2}}\hspace{0.1cm} ,
\end{eqnarray}
\begin{eqnarray}\label{eq:parameter3}
\rho_3=\sqrt{Q^2+\frac{\epsilon_3\omega_n^2}{c^2}}\hspace{0.1cm},
\end{eqnarray}
\begin{eqnarray}\label{eq:f}
\tilde f(\phi)=\frac{\sqrt{Q^2((\epsilon_{1,\parallel}/\epsilon_{1,\perp})-1)\cos^2\phi+\rho_{1,\perp}^2}-\rho_{1,\perp}}{Q^2 \sin^2\phi-\rho_{1,\perp}^2}\hspace{0.1cm} .\qquad
\end{eqnarray}
\begin{table}[b]
 \renewcommand{\arraystretch}{1.5}%change the table spacing
\caption{\label{modelparameter}
Model parameters used to determine the dielectric function of the materials\cite{dielectric2,dielectric3,Munday2}. }
\begin{ruledtabular}
\begin{tabular}{ccccc}
  & $C_{IR}$ & $C_{UV}$ & $\omega_{IR}(\rm rad/s)$ & $\omega_{UV}(\rm rad/s)$ \\
\hline
 Calcite$\parallel$ & $5.300$ & $1.683$ & $2.691\times10^{14}$ & $1.660\times10^{16}$ \\
    Calcite$\perp$ & $6.300$ & $1.182$ & $2.691\times10^{14}$ & $2.134\times10^{16}$ \\
    Barium titanate$\parallel$ & $3595$ & $4.128$ & $0.850\times10^{14}$ & $0.841\times10^{16}$ \\
    Barium titanate$\perp$ & $145.0$ & $4.064$ & $0.850\times10^{14}$ & $0.896\times10^{16}$ \\
    Silica & $0.829$ & $1.098$ & $0.867\times 10^{14}$ & $2.034\times 10^{16}$\\
\end{tabular}
\end{ruledtabular}
\end{table}

Here $\Delta_{\perp}=(\epsilon_{2,\perp}-\epsilon_{3})/(\epsilon_{2,\perp}+\epsilon_3)$, $\Delta_{\parallel}=(\epsilon_{2,\parallel}-\epsilon_3)/\epsilon_3$ are the relative anisotropy measures of the cylinder, $d$ is the separation between the cylinder and the half-space, $a$ is the radius of the cylinder, $k_B$ is Boltzmann constant, $T$ is temperature, $\epsilon_3$ is the dielectric response of the isotropic medium between the cylinder and the half space, $\epsilon_{1,\perp}$ and $\epsilon_{1,\parallel}$ are the dielectric responses of the birefringent material,  $\epsilon_{2,\perp}$ and $\epsilon_{2,\parallel}$ are the  dielectric responses of the cylinder material.
Subscript $n$ is the index for the Matsubara frequencies, which are $\omega_{n}=2n\pi k_{B}T/\hbar$, and the prime on the summation in Eq.~\ref{eq:freeenergy} means that the weight of the $n=0$ term is $1/2$. All the dielectric responses should be considered as functions of discrete imaginary Matsubara frequencies, i.e., as $\epsilon_3\equiv\epsilon_3^{(n)}=\epsilon_3(i\omega_n)$, $\epsilon_{1,\perp}(i\omega_n)$,  $\epsilon_{1,\parallel}(i\omega_n)$, $\epsilon_{2,\perp}(i\omega_n)$ and $\epsilon_{2,\parallel}(i\omega_n)$.

\begin{figure}[b]
\includegraphics[width=0.52\textwidth]{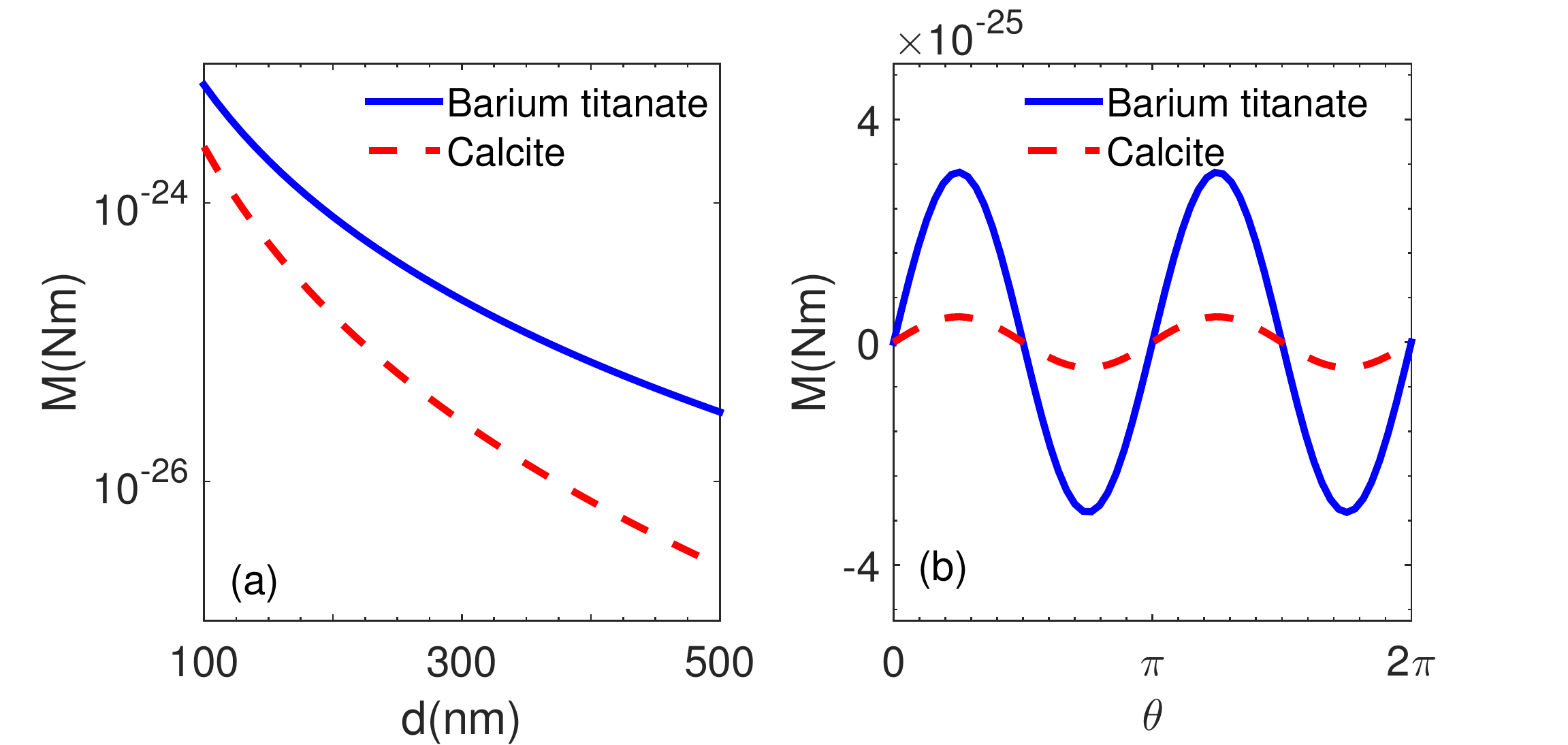}
  \caption{\label{torque}  (a) Calculated Casimir torque in 300K as a function of separation $d$ between the silica nanorod ($l=200~\rm nm,\ a=20~\rm nm$) and a birefringent crystal at relative orientation $\theta=\pi/4$. The blue solid line and red dashed line represent the torque when the birefringent crystals are barium titanate and calcite, respectively. (b) Calculated Casimir torque in 300K as a function of relative orientation $\theta$ between the silica nanorod  ($l=200~\rm nm,\ a=20~\rm nm$)and a birefringent crystal by a separation $d=266~\rm nm$. The blue solid line and red dashed line represent the torque when the birefringent crystals are barium titanate and calcite, respectively.}
\end{figure}

The dielectric properties of many materials are well described by a multiple oscillator model (the so-called Ninham-Parsegian representation)\cite{dielectric1}. For most inorganic materials, only two undamped oscillators are commonly used to describe the dielectric function\cite{dielectric2,dielectric3},
\begin{eqnarray}\label{eq:dielectric}
\epsilon(i\xi)=1+\frac{C_{IR}}{1+(\xi/\omega_{IR})^2}+\frac{C_{UV}}{1+(\xi/\omega_{UV})^2}\hspace{0.1cm},
\end{eqnarray}
where $\omega_{IR}$ and $\omega_{UV}$ are the characteristic absorption angular frequencies in the infrared and ultraviolet range, respectively, and $C_{IR}$ and $C_{UV}$ are the corresponding absorption strengths. For the birefringent materials, there are separate functions describing dielectric functions for the ordinary and extraordinary axis. The model parameters used for our calculations are summarized in Table~\ref{modelparameter}.

We could use Eq.~\ref{eq:freeenergy}-~\ref{eq:dielectric} and parameter data in Table~\ref{modelparameter} to calculate the Casimir free energy $G(d,\theta)=g(d,\theta)\times l$, where $l$ is the  length of the cylinder \cite{Mostepanenko}. FIG.~\ref{potentialandforce}.(a) inset shows that the Casimir free energy is very small for separation $d>100~\rm nm$, compared to the optical trapping potential. So the nanorod will be trapped near the center of the laser beam  without being attracted to the substrate by the Casimir force. When $d<100~\rm nm$ the size of the nanorod is comparable to the separation between the nanorod and the birefringent crystal, thus the dilute cylinder approximation will fail. Therefore, we only consider the situation when $d>100~\rm nm$.

Then the retarded Casimir (or Casimir-Lifshitz) force is given by
\begin{eqnarray}\label{eq:force}
F=-\frac{\partial G(d,\theta)}{\partial d}\hspace{0.1cm},
\end{eqnarray}
and the torque induced by the birefringent plates is given by \cite{Barash}
\begin{eqnarray}\label{eq:torque}
M=-\frac{\partial G(d,\theta)}{\partial \theta}\hspace{0.1cm}.
\end{eqnarray}

Using  Eq.~\ref{eq:force},~\ref{eq:torque}, we have calculated the Casimir force and torque expected for different separations at relative orientation $\theta=\pi/4$, both for the barium titanate and calcite as the birefringent crystal.
The results obtained for $T=300~\rm K$ are reported in FIG.~\ref{potentialandforce}.(b) and FIG.~\ref{torque}.(a).  The force and torque both decrease as the separation increases. The force follows the same power dependence of the separation for different birefringent materials. However, there is no single power law dependence that describes the torque at all separations regardless of the choice of materials. That is because the Casimir torque is directly determined by the dielectric response difference between ordinary and extraordinary axes, which is discrepant between barium titanate and calcite (in Table.1).

We have also calculated the Casimir torque at $d=266~\rm nm$ as a function of the relative orientation. From the results reported in FIG.~\ref{torque}.(b), one can clearly see that the torque oscillates sinusoidally  with periodicity of $\pi$: $M=M_0 \sin(2\theta)$. The maximum magnitude of the torque occurs at $\theta=\pi/4$ and $\theta=3\pi/4$. For different birefringent crystals, the maximum magnitudes of the torque are different, but have the same periodicity. As expected, materials with less birefringence give rise to a smaller torque. Thus Casimir torque between silica nanorod and calcite is smaller than that with barium titanate.

\section{Effects of thermal photons}
\begin{figure}
  \centering
    \includegraphics[width=0.52\textwidth]{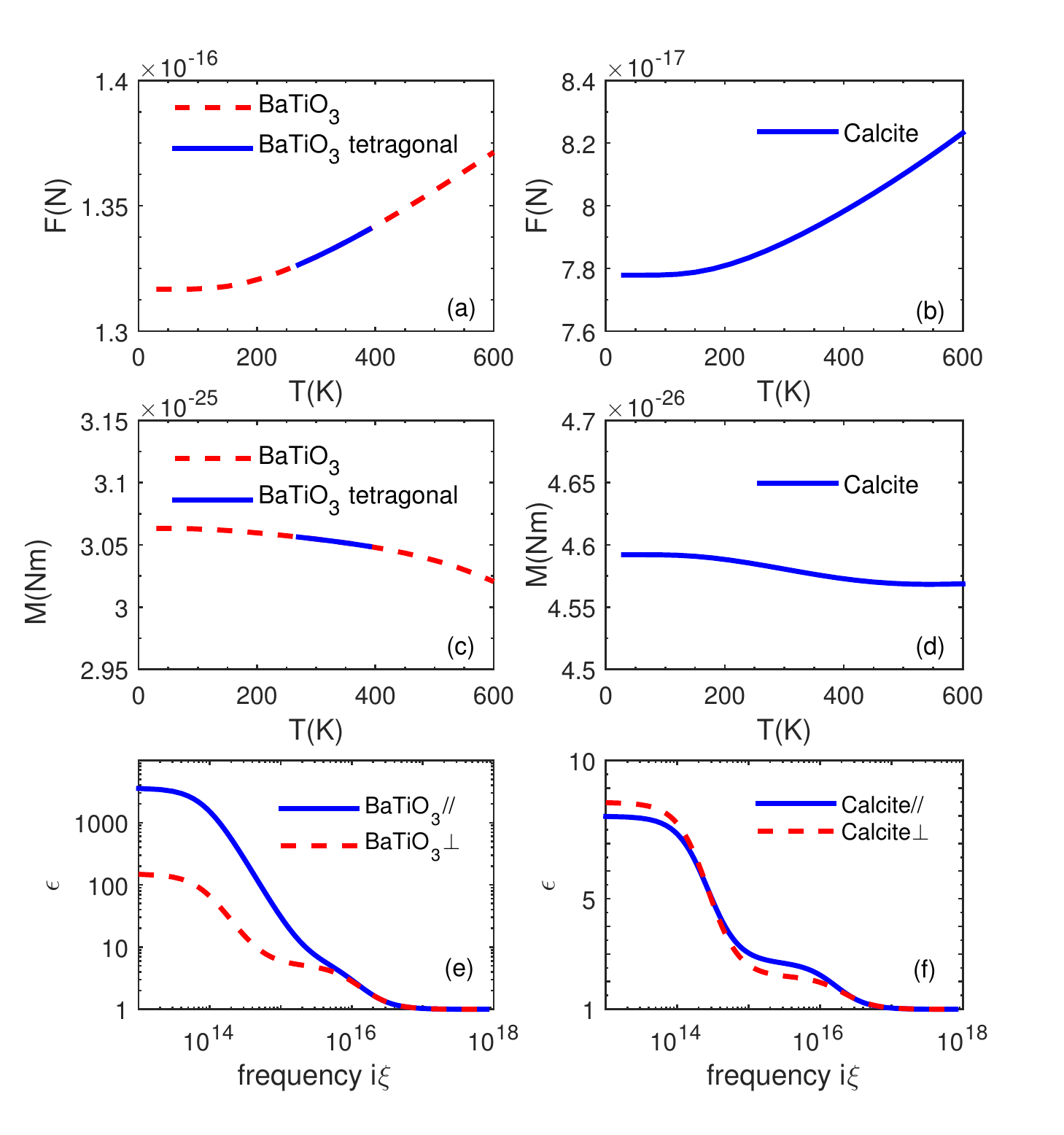}
      \caption{\label{temperature}(a),(b) Calculated temperature dependence of Casimir force between a silica nanorod and a birefringent crystal, separated with a distance of 266~nm and aligned with an angle of $\pi/4$. The materials of the birefringent crystals are barium titanate and calcite, respectively. To single out the effects of thermal photons, we did not consider the temperature dependence of dielectric parameters in the calculation. When the temperature is below 400K, the force is almost unchanged, with only 2 \% difference.  As the temperature increases, the casimir force increases and temperature dependence becomes more and more significant. For barium titanate, the temperature range for tetragonal structure is 278K-393K, the blue solid line represents this range. Calcite is a very stable birefringent crystal during the plotted temperature range. (c),(d) Calculated temperature dependence of Casimir torque. When the temperature is below 400K, the torque is almost unchanged. As the temperature increases, the casimir torque decreases and temperature dependence becomes more and more significant. (e),(f)Dielectric functions as a function of imaginary angular frequency for barium titanate and calcite. Blue solid lines and red dashed lines represent, $\epsilon_{\parallel}$ and $\epsilon_{\perp}$, respectively.}
\end{figure}

Several papers have reported measurements of the thermal Casimir force \cite{thermal1,thermal2}, which is due to thermal photons (blackbody radiation) at finite temperature rather than quantum vacuum fluctuations of the electromagnetic field. At room temperature, the thermal Casimir force is typically much smaller than the conventional Casimir force due to quantum vacuum fluctuations. The measurements of thermal Casimir force could test different models of materials. In fact, it is still under debate about how to calculate the thermal Casimir force between real materials \cite{thermal debate}.  It is thus interesting to see how thermal photons affect the Casimir torque and whether thermal Casimir torque will be detectable with our proposed method.

 When the separation between the nanorod and the birefringent crystal is relatively small and the temperature is relatively low, the blackbody radiation could be neglected. But when the temperature and the separation increase, a small fraction of the torque will come from thermal photons. To single out the effects of thermal photons, we assume the dielectric functions of materials are independent of temperature. The effect of temperature is only included in the Bose-Einstein distribution of thermal photons. In other words, we let the explicit temperature $T$ in Eq.~\ref{eq:freeenergy} and the Matsubara frequencies $\omega_n = 2 n \pi k_B T/ \hbar$  be a variable, while assuming all parameters listed in Table 1  to be constants. While this is a crude approximation, it can help us to understand the effects of thermal photons on the Casimir torque. In real experiments, the properties of materials will depend on temperature. So the situation will be more complex.

The calculated results of the Casimir force and torque for $d=266~\rm nm$ as a function of temperature are shown in FIG.~\ref{temperature}.(a)-(d). We can see that the thermal Casimir effect is very small (less than a few percent). So the measured torque will mainly come from quantum fluctuations. The purpose of this calculation is to estimate the magnitude of the effect of thermal photons. Since we did not consider the change of the dielectric constants of the real materials as a function of temperature, the temperature dependence of the experimental results is expected to be    different from FIG.~\ref{temperature}. To avoid complications, it will be better to do the experiment at a fixed temperature. Because the effect of thermal photons is very small for separations considered here, the measured temperature dependence of the Casimir torque will be most likely due to the temperature dependence of the dielectric functions, instead of  thermal photons.

\section{Torque measurement method}
For $d=266~\rm nm$, the maximum magnitude of the Casimir torque on a silica nanorod ($l=200~\rm nm,\ a=20~\rm nm$)  is around $3.2\times 10^{-25}~\rm Nm$ for barium titanate and around $4.6\times 10^{-26}~\rm Nm$ for calcite (Fig. \ref{torque}).
In order to prove that our optically levitated nanorod system is able to  detect the Casimir force and torque, we calculated the sensitivity of the force and the torque and the results will be shown in subsection A. and B.

In a real system, there are some other effects, such as stray fields, surface roughness and patch potential on the surface, which may introduce errors to the measurement. We will analyze these effects in subsection C. and D.

\subsection{Torque sensitivity}
To understand the limit of torque sensitivity in the quantum regime, one must consider the noise limit which comes from thermal fluctuations, as well as from photon recoil.
In air, the interaction between the nanorod and the thermal environment dominates the noise, thus the photon recoil from the laser can be neglected. However, in high vacuum, the dominant source of noise can come from the unavoidable photon recoil in the optical trap and sets an ultimate bound for the sensitivity.

For small oscillation amplitudes, the equation of motion of a harmonic torsional oscillator is
\begin{eqnarray}\label{eq:motion}
\ddot{\theta}+\gamma\dot{\theta}+\Omega_r^2\theta =M(t)/I\hspace{0.1cm} ,
\end{eqnarray}
where $\theta$ is the angular deflection of the oscillator, $\Omega_r$ is the frequency of rotational motion, $M$ is a fluctuating torque, $I$ is the moment of inertia around the torsion axis, and $\gamma$ is the damping rate of the torsional motion which can be written as $\gamma=\gamma_{th}+\gamma_{rad}$. Here $\gamma_{th}$ accounts for the interaction with the background gas, $\gamma_{rad}$ refers to the interaction with the radiation field.

When the torque fluctuation is from Brownian noise, the angular fluctuations of an oscillator excited by such a stochastic torque could be calculated. The thermal noise limited minimum torque that can be measured with a torsion balance is\cite{Haiberger}
\begin{eqnarray}\label{eq:thermaltorque}
	 M_{th}=\sqrt{\frac{4k_{B}TI\gamma_{th}}{\Delta t}} \hspace{0.1cm},
\end{eqnarray}
where $k_{B}$ is the Boltzmann constant, $T$ is the environment temperature,  and $\Delta t$ is the measurement time. The damping coefficient from thermal noise is $\gamma_{th}=f_r/I$.
$I=\frac{\rho\pi a^2l^3}{12}$ is the moment of inertia of the nanorod around its center and perpendicular to its axis, $\rho$ is the density of the nanorod, $a$ is the nanorod radius and $l$ is the nanorod length.
$f_r=k_BT/D_r$ is the rotational friction drag coefficient.
$D_r$ is the rotational diffusion coefficient for a rod in the free molecular regime and can be represented as \cite{MingdongLi}
\begin{eqnarray}\label{eq:Dr}
D_r=k_BTK_n/\{\pi\mu l^3\big[(\frac{1}{6}+\frac{1}{8\beta^3})\nonumber\\
+f(\frac{\pi-2}{48}+\frac{1}{8\beta}+\frac{1}{8\beta^2}+\frac{\pi-4}{8}\frac{1}{8\beta^3})\big] \}\hspace{0.1cm} ,
\end{eqnarray}
where $\beta=l/a$ is the rod aspect ratio, $K_n=\lambda/a$ is the Knudsen number, $\lambda=\frac{\mu}{p}\sqrt{\frac{\pi k_BT}{2m_{gas}}}$ is the mean free path, $m_{gas}$ is the molecular mass, $ \mu$ is the gas viscosity and $f$ is the momentum accommodation, where we choose $f=0.9$.
Thus the minimum detectable torque due to thermal fluctuations $M_{th}$ decreases with the square root of the measurement duration $\Delta t$, while increases with the square root of the pressure $p$.

\begin{figure}
  \centering
\includegraphics[width=0.49\textwidth]{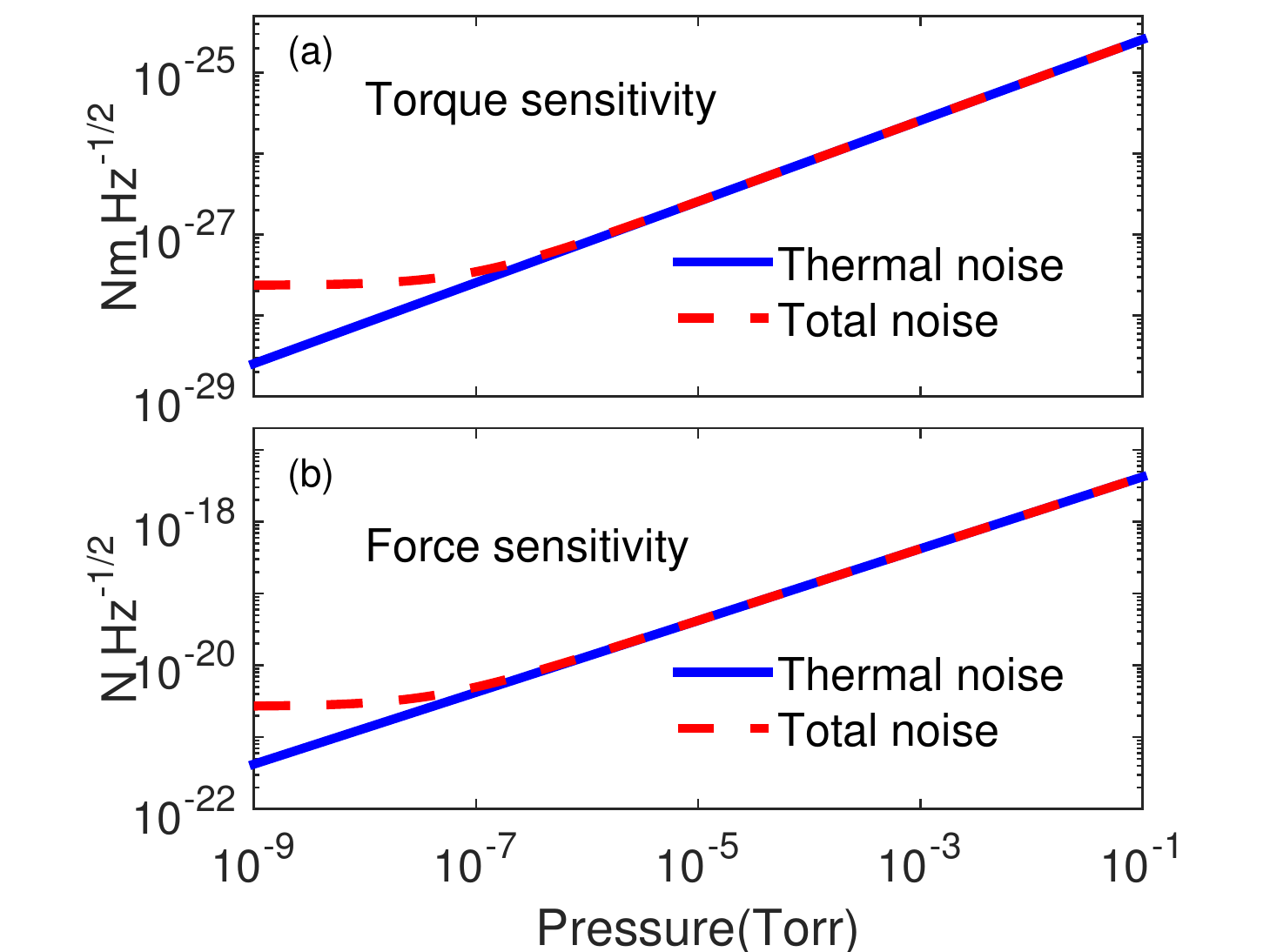}
\caption{\label{noise}(a)Calculated torque sensitivity limit of the rotational motion of a levitated silica nanorod ($l=200$ nm, $a=20$ nm) as a function of the background  gas pressure. The blue solid line and the red dashed line represent the torque sensitivity limit due to only the thermal noise and due to both photon recoil and thermal noise. The laser power is 100 mW and the laser waist radius is 400 nm. (b)Calculated force sensitivity limit of the translational motion of a levitated Silica nanorod as a function of gas pressure. The blue solid line and the red dashed line represent the force sensitivity limit due to only the thermal noise and due to both photon recoil and thermal noise.}
\end{figure}

Apart from fluctuations due to contact with the background gas, the unavoidable photon recoil from the optical trap also contributes to the noise limit.
The shot noise due to photon recoil can be understood as momentum or angular momentum kicks from the scattered photons.
The photon recoil limited minimum torque is
\begin{eqnarray}\label{eq:shottorque}
	 M_{rad}=\sqrt{\frac{4I}{\Delta t}\frac{d}{dt}\overline{K_R}}\hspace{0.1cm},
\end{eqnarray}
where rotational shot noise heating rate of a nanorod from a linearly polarized trapping beam is \cite{Changchun, Changchun2, decoherence}
\begin{eqnarray}\label{eq:heatingratetorque}
	\frac{d}{dt}\overline{K_{R}} = \frac{8\pi J_p}{3}\left(\frac{k_0^2}{4\pi\epsilon_0}\right)^2 (\alpha_{\perp}-\alpha_{\parallel})^2\frac{\hbar^2}{2I}\hspace{0.1cm}.
\end{eqnarray}
Here the photon flux $J_p$ is equal to the laser intensity over the energy of a photon, which means $J_p=I_{laser}/\hbar\omega_0$. $\omega_0$ is the frequency of incident beam, $k_0$ is the incoming wave vector.
Therefore, the total torque limit is given by
\begin{eqnarray}\label{eq:totaltorque}
	 M_{min}=\sqrt{M_{th}^2+M_{rad}^2}\hspace{0.1cm}\hspace{0.1cm},
\end{eqnarray}

We calculate the torque limit by using Eq.~\ref{eq:thermaltorque} -~\ref{eq:totaltorque} and the result of the calculation is shown in FIG.~\ref{noise}.(a). The torque detection sensitivity of a levitated nanorod will be limited by the thermal noise when the pressure is above $10^{-7}$ torr. When the pressure is below $10^{-7}$ torr, the torque sensitivity is mainly limited by photon recoil from the 100 mW trapping laser, and  is around $10^{-28}~\rm Nm/\sqrt{\rm Hz}$. Thus the Casimir torque will be 3 orders larger than the minimum torque our system can detect in 1 second, and is expected to be measurable.

\subsection{Force sensitivity}
 Similar to the torque sensitivity,  both thermal noise and photon recoil needs to be considered to determine force sensitivity. For small oscillation amplitudes, the nanorod's motion is described by three independent harmonic oscillators (for three directions), each with its own oscillation frequency $\Omega_{0i}$ and damping rate $\gamma_i$, which is a result of the asymmetric shape of the optical potential. For example, the motion along y is described by,
\begin{eqnarray}\label{eq:motion2}
\ddot{y}+\gamma_y\dot{y}+\Omega_{0y} ^2 y=\frac{1}{m}F_y(t)\hspace{0.1cm},
\end{eqnarray}
where $y$ is the motion of the center of mass, $m$ is the nanorod mass, $F_y$ is a fluctuating force along y axis acting on the nanorod. The thermal noise limited minimum force in one direction $i$ that can be measured with a force balance is
\begin{eqnarray}\label{eq:thermalforce}
	F_{th(i)}=\sqrt{\frac{4k_{B}Tm\gamma_{i}}{\Delta t}}\hspace{0.1cm},
\end{eqnarray}
Here m is the mass, $\gamma_{i}$ is the damping coefficient of the translational motion due to the background gas. For a nanorod, damping coefficients are directly related to the drag coefficients at each directions, which means that $\gamma_{\perp}=K_{\perp}/m$ (component perpendicular to the axial direction) and $\gamma_{\parallel}=K_{\parallel}/m$ (component parallel to the axial direction).
In the free molecular regime, the drag force along different directions for a cylindrical particle are expressed by $F_{\perp}=K_{\perp}V_{\perp}$ and $F_{\parallel}=K_{\parallel}V_{\parallel}$ and drag coefficients are\cite{MingdongLi2}
\begin{eqnarray}\label{eq:Kpara}
	K_{\perp}=\frac{2\pi\mu a^2}{\lambda}\left[(\frac{\pi-2}{4}\beta+\frac{1}{2})f+2\beta\right]\hspace{0.1cm},
\end{eqnarray}
\begin{eqnarray}\label{eq:Kperp}
	K_{\parallel}=	\frac{2\pi\mu a^2}{\lambda}\left[(\beta+\frac{\pi}{4}-1)f+2\right]\hspace{0.1cm},
\end{eqnarray}
	
	In our system, we only consider the motion perpendicular to the axis, which will affect the measurement of Casimir force. The force inducted by thermal fluctuations is
\begin{eqnarray}\label{eq:thermalforce2}
F_{th}=\sqrt{\frac{4k_{B}T}{\Delta t}K_{\perp}},
\end{eqnarray}
while the photon recoil limited minimum force is
\begin{eqnarray}\label{eq:shotforce}
F_{rad}=\sqrt{\frac{4m}{\Delta t}\frac{d}{dt}\overline{K_{T}}}\hspace{0.1cm}.
\end{eqnarray}
Here the translational shot noise heating rate of a nanorod from a linearly polarized trapping beam is\cite{Changchun}
\begin{eqnarray}\label{eq:heatingrateforce}
\frac{d\overline{K_{T}}}{dt}=\frac{8\pi J_{p}}{3}\left(\frac{k_0 ^2}{4\pi\epsilon_0}\right)^2\alpha_{\perp}^2\frac{\hbar ^2 k_{0}^2}{2m}\hspace{0.1cm},
\end{eqnarray}
Therefore, the total force limit will be
\begin{eqnarray}\label{eq:totalforce}
	 F_{min}=\sqrt{F_{th}^2+F_{rad}^2}\hspace{0.2cm}\hspace{0.1cm},
\end{eqnarray}
Then we use Eq.~\ref{eq:Kperp}-~\ref{eq:totalforce} to calculate the force sensitivity limit and the result is shown in FIG.~\ref{noise}.(b). The force detection sensitivity will be limited by the thermal noise when the pressure is above $10^{-7}$ torr. When the pressure is below $10^{-7}$ torr, the force sensitivity is mainly limited by photon recoil, which is about $10^{-21}~\rm N/\sqrt{\rm Hz}$. The Casimir force is approximately $10^{-16}~\rm N$ at $d=266~\rm nm$, therefore, it is expected to be measurable.

\subsection{Pulsed measurement scheme}

 Since the reflectances  along the ordinary and extraordinary axes of the birefringent plate are different, the reflected light  will not have the same polarization as the incident light (Fig. \ref{setup}). Thus there will be an optical torque from the laser reflected by the birefringent plate. To eliminate this effect, we will apply a pulsed measurement scheme, which means to switch the optical tweezer on and off repeatedly to detect the Casimir torque by observing the rotation of the nanorod. We could extract the contribution which  comes from the Casimir effect to the rotation during the period when the laser is off.

\begin{figure}
  \centering
\includegraphics[width=0.5\textwidth]{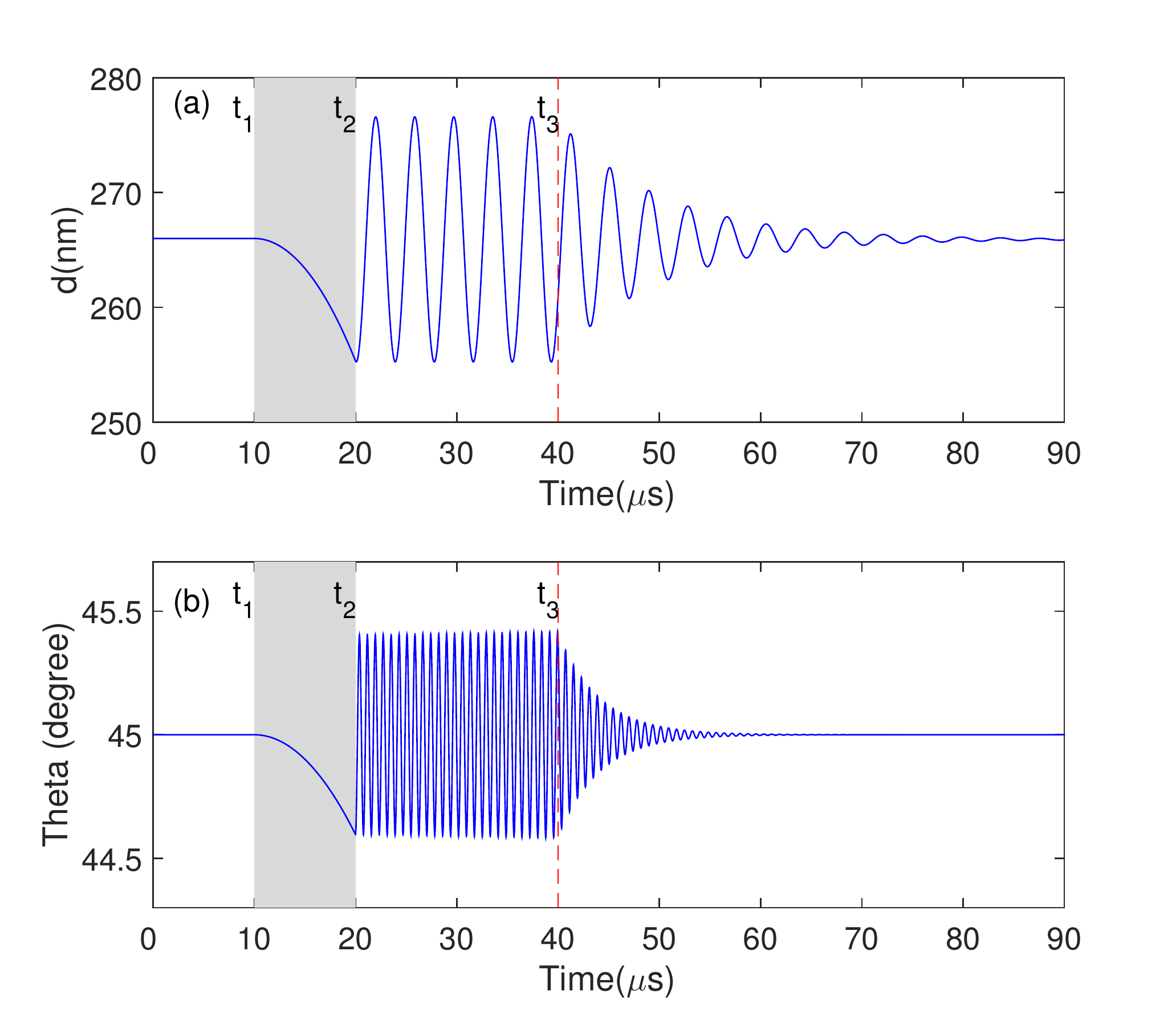}
\caption{\label{evolution}(a) Separation evolution diagram. (b) Relative orientation diagram.
$t=0$ to $t_1=10 \mu s$, the laser is on. $t_1=10 \mu s$ to $t_2=20 \mu s$, the laser is off. $t_2=20 \mu s$ to $t_3=40 \mu s$, the laser is on again. During this period, there is no feedback cooling. $t_3=40\mu s$, the laser is still on and we add feedback cooling to the nanorod.
}
\end{figure}

 When the laser is off, the naonorod will experience the torque  attributed to the Casimir effect. We can use this method to extract the torque from the Casimir part. However, when the laser is off, the nanorod will fall to the substrate by gravity as well as the Casimir force between it and the substrate. Therefore, we may lose the nanorod when the off-period is too long, while increasing the length of the off-period can amplify the signal observed from the Casimir torque.
 FIG.~\ref{evolution} shows the simulation for this method. Initially the polarization of the laser is set to be $45^{\circ}$ relative to the optical axis of the birefringent plate, and the center of the laser beam is set at a distance of 266 nm from the substrate. FIG.~\ref{evolution}.(a) is the separation evolution during the pulse measurement. During the period 0 to $t_1$ we will keep the laser on. During this time, the nanorod is trapped stably around the center of the beam ($d=266$nm), which is an equilibrium position. At $t=t_1=10\mu s$, we will turn off the laser. So at this moment the nanorod will fall to the substrate with a acceleration of about $200 m/s^2$ due to gravity and the Casimir force. We notice that, the nanorod will only fall 10 nm for a 10 $\mu s$ period. At $t=t_2=20\mu s$, we turn on the laser again. The trapping force from the laser will pull back the nanorod. Then the nanorod will do harmonic oscillations around the equilibrium position ($d=266$ nm). When time reaches $t_3=40 \mu s$, we will intentionally apply feedback cooling
 to the nanorod \cite{feedbackcooling, forcesql}. So the amplitude of the nanorod will decay to almost zero. At the end of a measurement cycle, the nanorod will come back to the initial situation.
FIG.~\ref{evolution}.(b) is the angle evolution during a pulsed measurement cycle. When the laser is on, the nanorod experiences an optical torque from the incident laser beam and from the reflected light, as well as a relatively small Casimir torque. Since the optical torque provided by the trapping laser can be far larger than the  Casimir torque, the initial relative orientation is approximately $45^{\circ}$. When the laser is off during $t_1=10 \mu s$ and $t_2=20 \mu s$, the torque  comes only from the Casimir effect. We will repeat this sequence many times (could be millions of times \cite{Hoang2017}) and average the results to extract the signal from the noise. The effective measurement time will be $t_2-t_1$ times the number of measurement cycles.

\subsection{Other effects}

In real experiments,  there could be external stray fields, surface roughness and surface patch potentials that could affect measurements.
If the nanorod has a permanent electric or magnetic dipole, there may be a torque on the nanorod due to stray electric  or magnetic fields. Different from the Casimir torque, such torque due to a permanent dipole has a period of $2\pi$. So if we rotate the nanorod by $180^{\circ}$, the dipole torque will change its sign, while the Casimir torque will be the same (FIG.~\ref{torque}(b)). Thus we can cancel out this dipole torque on the nanorod by a careful design.

\begin{figure*}
  \includegraphics[width=1\textwidth]{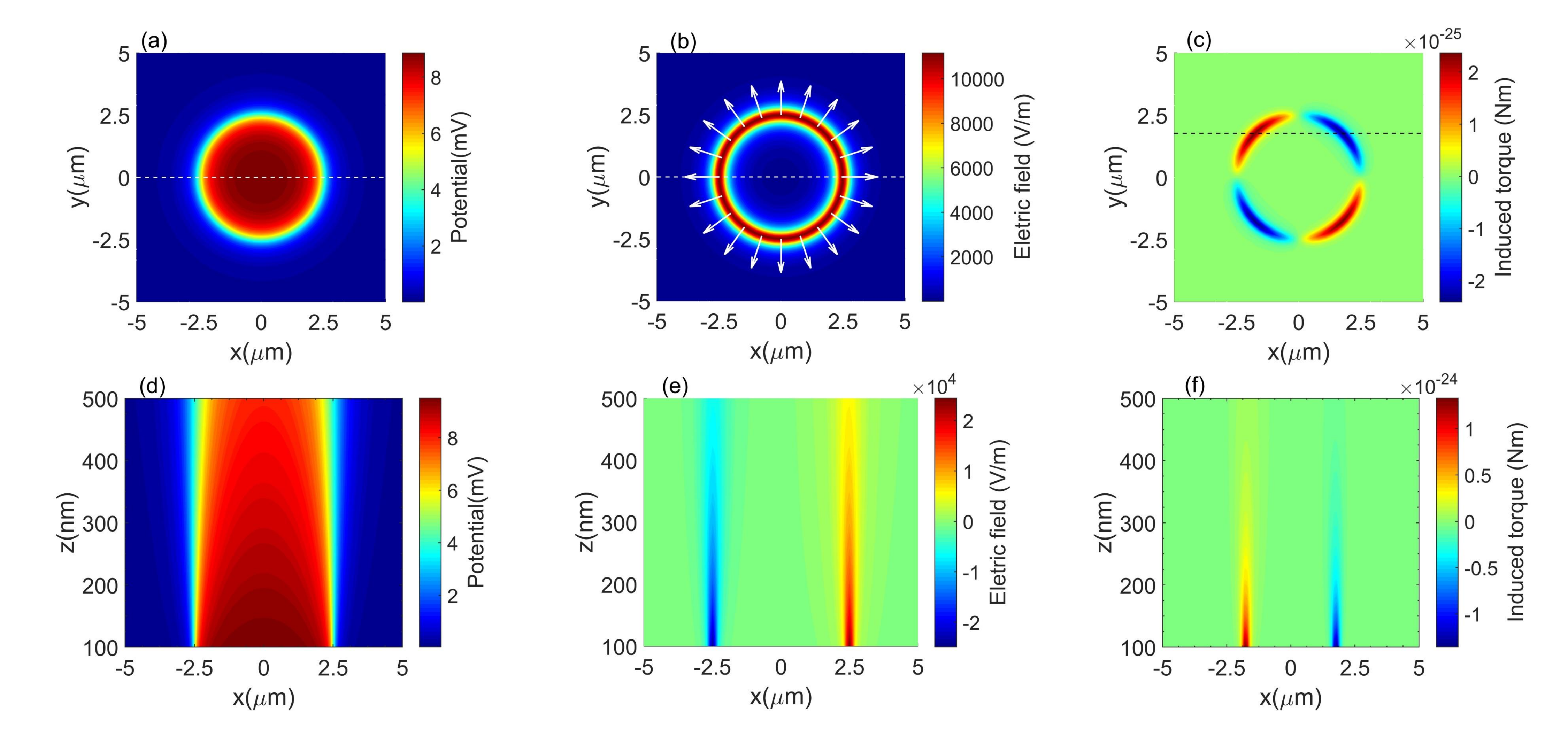}
  \caption{\label{patch} Calculation results of the electric potential, electric field and induced torque on the nanorod due to a patch potential. We assume that a round patch with a diameter of 5 $\mu$m lies on the birefringent plate and it is maintained at a potential of 10 mV, while the plate is kept at zero potential. (a) The potential at a height of 266 nm above the plate (a plane parallel to the birefringent plate). The center of the round patch is at (0,0). (b) The amplitude of the electric field at the height of 266 nm above the plate. The electric field reaches its maximum at the edge of the patch and is always pointing away from the center of the patch. The white arrows show the directions of the electric field. (c) The  torque induced by the patch potential. The axis of the nanorod is parallel to the $y$-axis. (d) The potential in a plane perpendicular to the birefringent plate and going through the center of the patch ($y=0$, shown as the white dashed line in (a)). The center of the patch is at (0,0), which is not shown in the plot. (e) The electric field from the patch in the X-O-Z plane (shown as the white dashed line in (b)). Here we only consider the electric field along x axis. (f) The torque induced by the patch potential in a plane perpendicular to the birefringent plate and 1.77 $\mu$m from the center of the patch ($y=1.77 \mu m$, shown as the black dashed line in (c)). }
\end{figure*}

\begin{figure}[t]
  \includegraphics[width=0.48\textwidth]{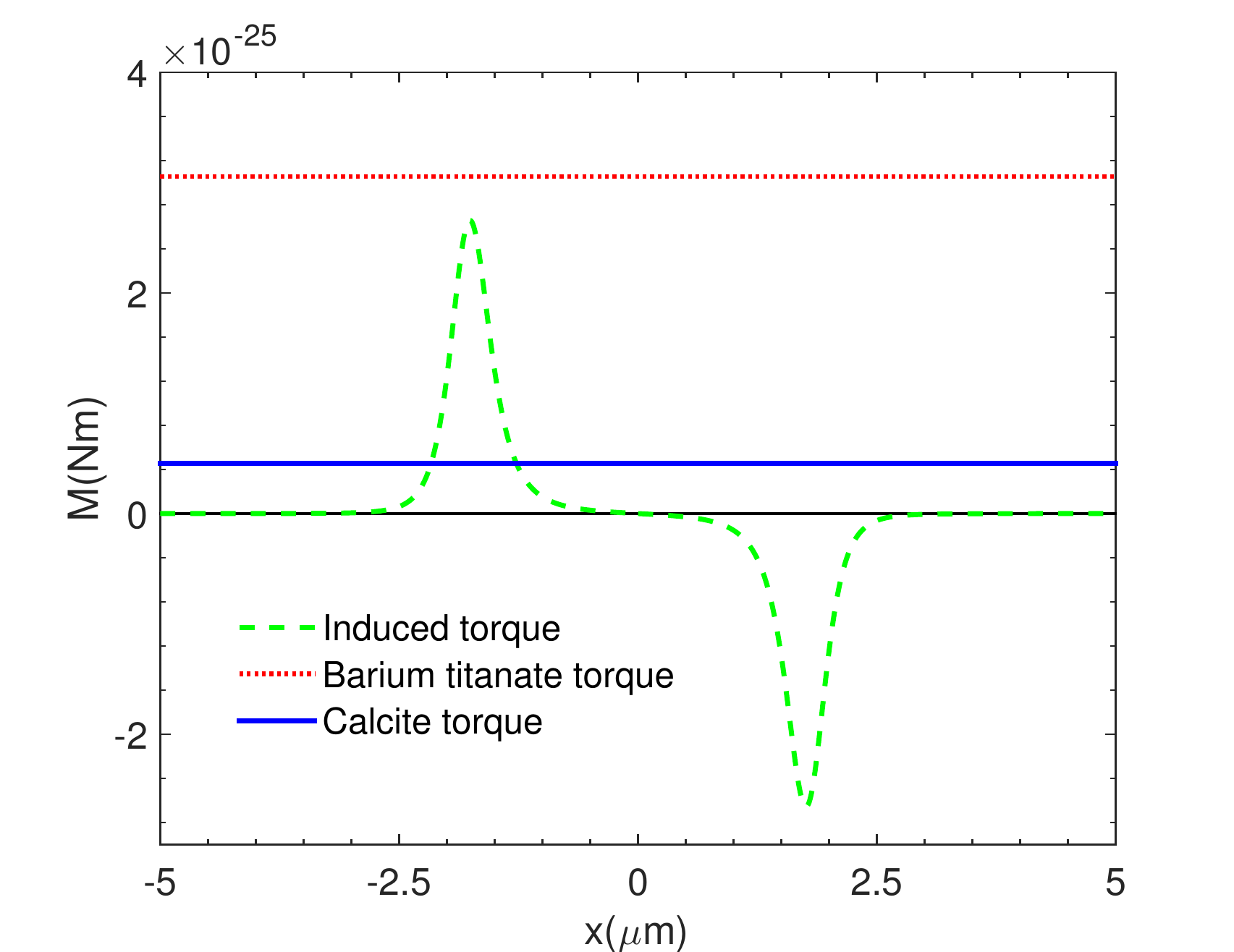}
  \caption{\label{profile} The torque profile at $z=266$ nm, $y=1.77 \mu m$ when there is a round patch with a diameter of 5 $\mu$m at the center. The patch is maintained at a potential of 10 mV. The green dashed line is the induced torque from the patch potential on the surface (corresponding to a horizontal line at $z=266$ nm in FIG.~\ref{patch}(f)). The red dotted line is the Casimir torque between a silica nanorod and a barium titanate plate at a separation of 266 nm (the same result as FIG.~\ref{torque}). The blue solid line is the Casimir torque between a silica nanorod and a calcite plate at a separation of 266 nm. The maximum value of the induced torque is at the same order as the Casimir torque, for both barium titanate and calcite situation. But the induced torque will change its sign at different positions. Therefore, we can cancel out the effect from patch potential by averaging the measured torque at multiple locations.}
\end{figure}

Roughness can change the effective separation between the nanorod and the plate, and  induce an additional torque on the nanorod. However, after  polishing, the roughness could be controlled to be less than 3 nm for the regime near the nanorod \cite{surfaceeffect}. FIG.~\ref{potentialandforce}(b) and FIG.~\ref{torque}(a) show the dependence of Casimir force and torque on the separation. When the separation changes 3 nm at an average separation about 266 nm, the force and the torque don't change  much. Therefore, the torque generated by surface roughness can be neglected.

Besides, there could be an inhomogeneous surface patch potential along the surface of real materials.
Such patch potential can introduce a force and a torque, which can affect Casimir force and torque measurements. Luckily, it has been experimentally demonstrated that most  levitated nanoparticles have zero electric charge, which can be verified by driving the particle with an AC electric field\cite{zeptonewtonforce, netcharge}. 

When there is no charge on the nanorod, the electric field of a patch potential can still cause a force and a torque due to the induced dipole. Here we analyze this situation and consider that there is a round  patch with a diameter of 5 $\mu$m on a large birefringent plate. The patch is assumed to have a potential of 10~mV,  while the plate is kept at zero potential.
Then the potential at any point above the plane is given by \cite{Jackson}\cite{Surfacetrap}
\begin{eqnarray}\label{eq:patchpotential}
\Phi(\rho,z)=V_{0}r_{0}\int_0^{\infty}e^{-\lambda z}J_{1}(\lambda r_{0})J_{0}(\lambda\rho)d\lambda,
\end{eqnarray}
where $r_0=2.5 \mu$m is the radius of the patch, $V_{0}=10$ mV is the fixed potential of the patch, $J_0$ and $J_1$ are the zero-order and first-order Bessel function of the first kind,  $\rho$ and z are the position of the potential in polar coordinates. 
  The long axis of the rod aligns with the laser polarization, which is assumed to be y direction in FIG.~\ref{patch}.(a). Then the induced torque along z axis from the patch potential on the nanorod becomes
\begin{eqnarray}\label{eq:patchpotential}
T_z=(\alpha_{\perp}-\alpha_{\parallel})E_{x}E_{y},
\end{eqnarray} 
where $\alpha_{\parallel}$ and $\alpha_{\perp}$ are the components of the DC polarizability tensor parallel and perpendicular to the  axis of the nanorod, $E_{x}$ and $E_{y}$ are the electric field along x axis and y axis, respectively. Here the positive torque is in the direction along the positive z axis.
 
 The calculated results of the electric potential, electric field and induced torque on the nanorod due to the patch potential are shown in FIG.~\ref{patch}. FIG.~\ref{patch}(a) is the calculated patch potential in the plane 266 nm above the birefringent plate. FIG.~\ref{patch}.(b) is the calculated electric field in the plane 266 nm above the birefringent plate. We can see that the electric field reaches the maximum value at the edge of the patch and is always pointing away from the center of the patch. FIG.~\ref{patch}.(c) shows the induced torque from the patch potential and it also reaches the maximum at the edge. The torque at the edge could reach $2\times 10^{-25}$ Nm, which is at the same order as the Casimir torque. 
FIG.~\ref{patch}.(d) is the calculated potential in a plane perpendicular to the birefringent plate (y=0, shown as the white dashed line in (a)). Here the height above the plate ranges from 100 nm to 500 nm. FIG.~\ref{patch}.(e) is the calculated patch potential in the X-O-Z plane (y=0, shown as the white dashed line in (b)).  Here we only consider the electric field along x axis.  We can see that the electric field  is anti-symmetric with respect to the z axis and is large at the edge of the patch.  FIG.~\ref{patch}.(f) shows the induced torque from the patch potential in a plane perpendicular to the plate but has a distance of 1.77 $\mu$m from the center of the patch ($y=1.77 \mu m$, shown as the black dashed line in (d)). In this plane, the maximum electric field is $45^{\circ}$ relative to the plane.  The torque also reaches the maximum at the edge. 

FIG.~\ref{profile} provides a more detailed profile of the induced torque at a height of 266 nm and 1.77 $\mu$m from the center of the patch (corresponding to a horizontal line in FIG.~\ref{patch}(f)). Comparing the induced torque by the patch potential with the Casimir torque between silica nanaorod and two birefringent plates, we can see that the maximum value of the induced torque is at the same order as the Casimir torque. However, when the nanorod is not close to the edge of the patch, the induced torque is far smaller than the Casimir torque.
Besides, the induced torque has different signs at different positions, while the Casimir torque is independent of the location for a single crystal birefringent plate. Therefore, we can cancel out the torque from the patch potential by measuring the torque at multiple locations along a line (1D) or along a 2D array. 

\begin{figure*}[t p]
  \includegraphics[width=1.05\textwidth]{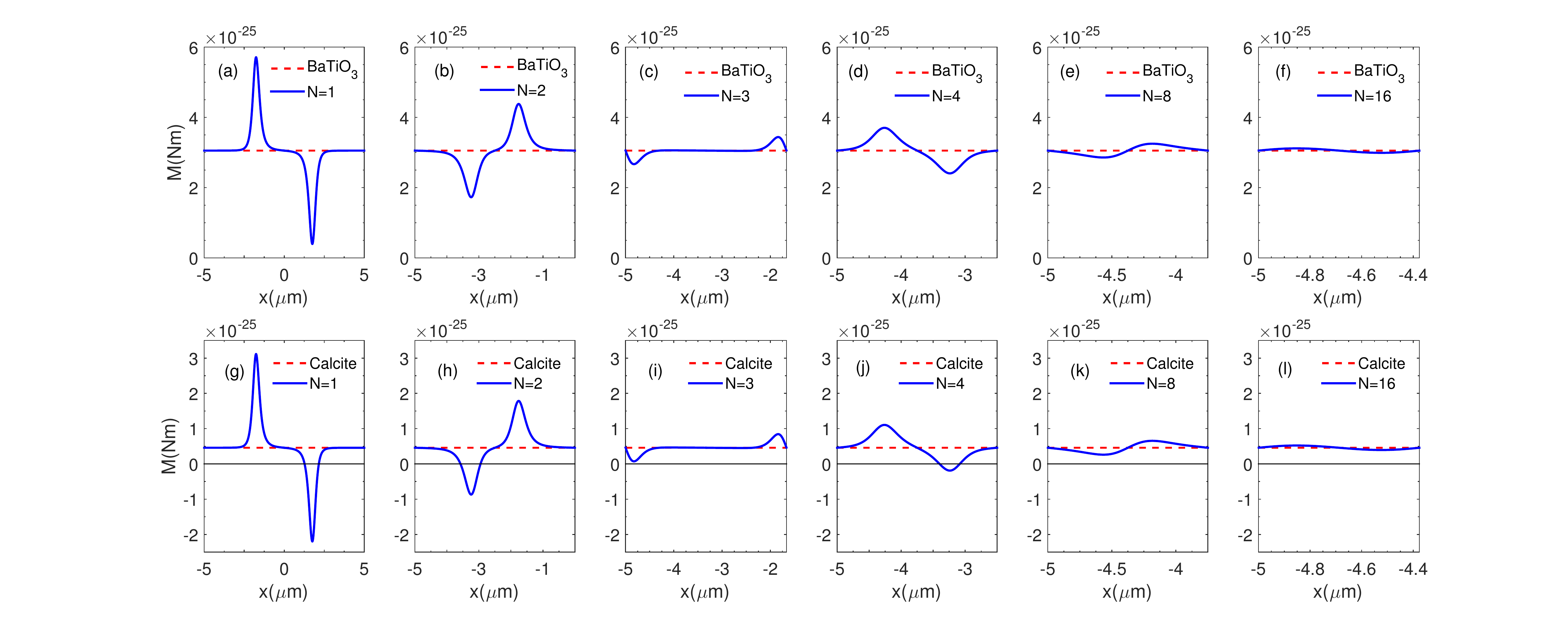}
  \caption{\label{equalmeasure} The torque after averaging the measured torque at equally spaced positions. We assume there is a round patch with a diameter of 5 $\mu$m at the center ($x=0$, $y=0$, $z=0$) and the patch is maintained at a potential of 10 mV. We compare the patch-induced torque at $z=266$ nm, $y=1.77 \mu$m (corresponding to a horizontal line in FIG.~\ref{patch}(f)) with the Casimir torque between the silica nanorod and the birefringent plates (barium titanate and calcite). (a) The blue solid line shows the expected measured torque when the patch-induced torque is included,  which means $M=M_{Casimir}+M_{patch}$. The red dashed line is the Casimir torque between the nanorod and barium titante plate. (b)-(f) show the torque after averaging the measured torque at $N$ different positions and these positions are equally spaced on the plate. We can see that the torque from the patch decays very fast when we increase $N$. (g)-(l) show  similar results when the birefringent plate is a calcite plate.}
\end{figure*}

\begin{figure}[t p]
  \includegraphics[width=0.5\textwidth]{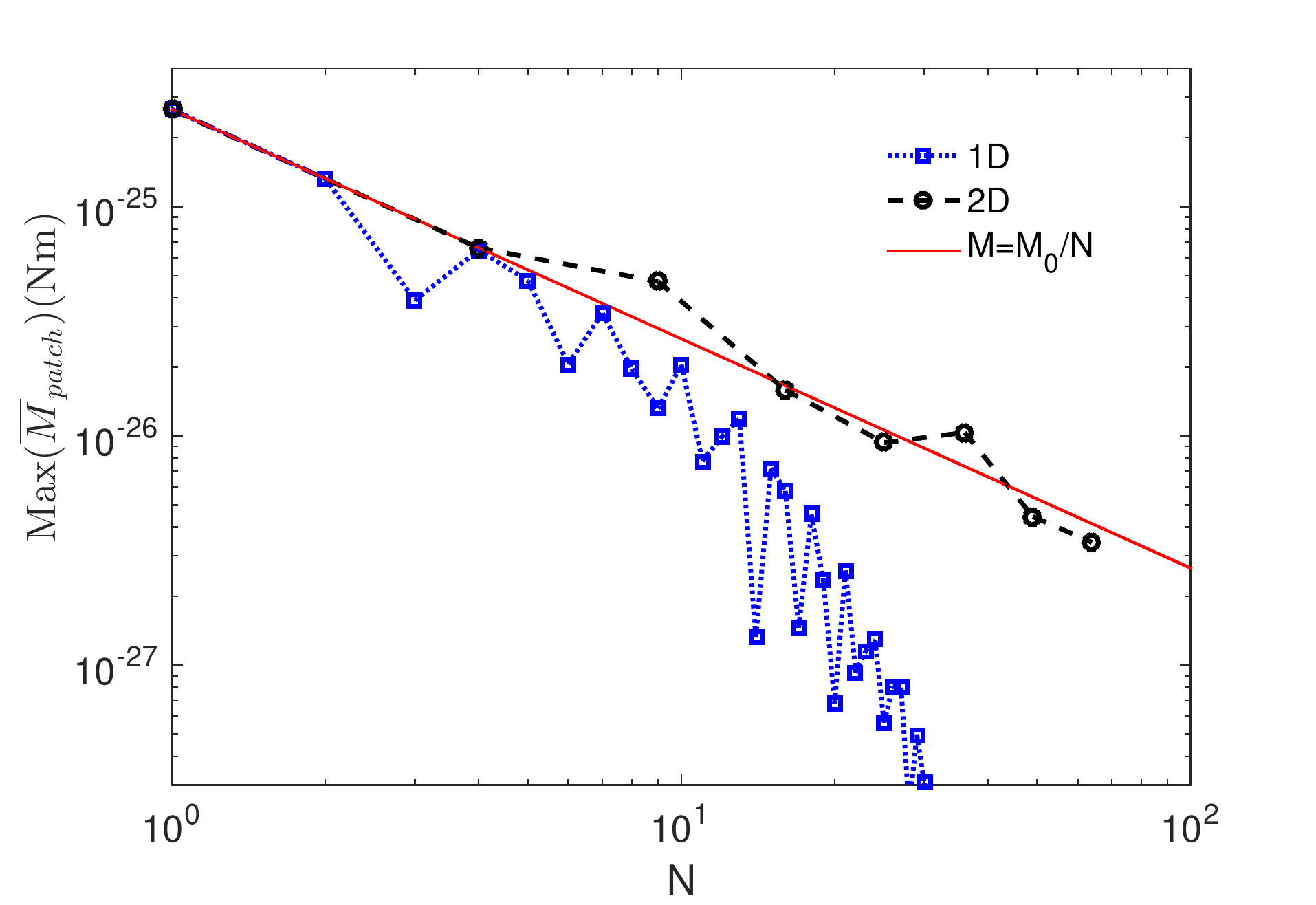}
  \caption{\label{decayN} The maximum value of the averaged torque from a patch potential $Max(\overline{M}_{patch})$ as a function of the total measurement number $N$. The effect from the patch potential decreases after averaging the measured torque at equally spaced positions.   Here the blue dotted line shows the maximum torque after averaging for one-dimension measurements along the $x$ axis, while the black dashed line shows the maximum torque after averaging for two-dimension measurements. The red solid line shows the equation $M=M_0/N$, where $M_0=2.65\times 10^{-25}$Nm is the maximum value of the patch-induced torque at 266 nm above the birefringent plate (no averaging). }
\end{figure}

One possible way is to measure the torque at points that are equally spaced on the birefringent plate. 
For 1D scan, we will measure the points along the direction which is perpendicular to the axis of the nanorod. We assume that the axis of the nanorod is trapped along y direction, then we will measure the torque along positions that are equally spaced along x axis. In this way, the average of measured torque will be 
\begin{eqnarray}\label{eq:equalmeasure}
\overline{M}_{1D}(x)=\frac{1}{N}\sum_{i=0}^{N-1}M(x+\frac{L}{N}i),
\end{eqnarray} 
where $x\in(0,\frac{L}{N})$ is the position of the first measurement point, $N$ is the total number of  measurements, $L$ is the measurement range, and $M(x+\frac{L}{N}i)$ is the torque measured at the $i$-th point, which includes both Casimir torque and the torque from patch potential. We let $x$ be a variable to simulate the situation when we do not know the exact location of the patch. The measurement range $L$ should be larger than the size of a patch. In the situation when there are many patches, $L$ will be the larger the better for a single crystal birefringent plate. 
Here we simulate the results for the situation we discussed before in FIG.~\ref{profile} and set $N$ to be 1, 2, 3, 4, 8 and 16. The results are shown in FIG.~\ref{equalmeasure}. We compare the effect from the patch at $z=266$ nm, $y=1.77$ $\mu$m with the Casimir torque between the silica nanorod and the birefringent plates (barium titanate and calcite). Here we set the measurement range $L=10$ $\mu$m. FIG.~\ref{equalmeasure}.(a) shows the expected measured torque (blue solid line) when the patch-induced torque is included, which means $M=M_{Casimir}+M_{patch}$. The red dashed line shows the Casimir torque between the nanorod and barium titanate plate. (b)-(f) show the average of the torque measured at $N$ equally spaced positions on the plate when $N=$2, 3, 4, 8, 16, respectively. We use Eq.~\ref{eq:equalmeasure} to the get the average torque. We can see that the average torque from the patch decays very fast when we increase $N$. FIG.~\ref{equalmeasure}(g)-(l) show  similar results when the birefringent plate is a calcite plate. In real experiments, we may not know the position of the patch. However, as shown in Fig. \ref{equalmeasure}(f),(l),  when N is large, the torque from the patch potential will be negligible compared to the Casimir torque and will only have a weak dependence of the measurement position.
 Therefore, we can use this method to minimize the effect from patch potential.

We also consider a two-dimensional scan, which means we will measure the torque following a 2D array in the X-O-Y plane.
In this way, the averaged torque will be 
\begin{eqnarray}\label{eq:equalmeasure2d}
\overline{M}_{2D}(x,y)
=\frac{1}{K^2}\sum_{i=0}^{K-1}\sum_{j=0}^{K-1}M(x+\frac{L}{K}i,y+\frac{L}{K}j),
\end{eqnarray} 
where $(x,y)$ is the position of the first measurement point on the plate and $x\in(0,\frac{L}{K}),y\in(0,\frac{L}{K})$. $K$ is the number of measurements along one axis, $N=K^2$ is the total number of  measurements, $L$ is the measurement range in one dimension and $M(x+\frac{L}{K}i,y+\frac{L}{K}j)$ is the torque measured at the $i$-th point along x axis and the $j$-th point along y axis. Here we also choose measurement range $L$ to be $10$ $\mu$m. FIG.~\ref{decayN} shows the relation between the number of measurements and the averaged torque from patch potential, both for one-dimensional (blue dotted line) and two-dimensional scans (black dashed line). Vertical axis shows the maximum of the averaged torque from the patch potential, horizontal axis corresponds to the number of measurements. The red solid line shows  $M=M_0/N$, where $M_0=2.65\times 10^{-25}$Nm is the maximum value of the patch-induced torque at 266 nm above the birefringent plate (no averaging). For the two-dimensional averaging method, the maximum averaged torque from the patch potential approximately follows the 1/N law. The 1D averaged torque from the patch along $x$ axis decays much faster than the 2D averaged torque from the patch potential. When $N=30$, the maximum value of the 1D averaged torque from the patch potential is about $3\times 10^{-28}$ Nm, which is three orders smaller than the Casimir torque between a silica nanorod and a barium titanate plate. This further proves that we can decrease the effect from patch potential by measuring the torque at multiple locations.

%% 1d is better??

 Meanwhile, we can reduce the surface patch potential by careful preparation of the sample, as done in an experiment that measured the Casimir force which improved the flatness to be less than 3 nm over mm$^2$ area\cite{CasimirTang}.  We can determine the topography and observe patch potential on the surface by using Kelvin probe force microscopy and choose the area with relatively small roughness and patch potential to do the measurement\cite{surfaceeffect, KPFM}. We can also measure the toque due to the surface patch potential directly by utilizing the angular dependence of the Casimir torque. As shown in FIG.~\ref{torque}.(b), the Casimir torque will be maximum when the relative angle between the nanorod and the optical axis of a birefringent crystal is 45$^\circ$, and it will be 0 when the angle is 0$^\circ$ or 90$^\circ$. So we can  directly measure the the torque due to the patch potential by setting the angle to be 0$^\circ$  and 90$^\circ$  when the Casimir torque is zero. We can then subtract the torque due to the surface patch potential from the total measured torque to obtain the Casimir torque at 45$^\circ$.

%\vspace{0.3cm}

\section{Conclusion}
In this paper, we show that the calculated Casimir force  is on the order of $10^{-16}~\rm N$ and the torque is on the order of $10^{-25}~\rm Nm$ between an optically levitated silica nanorod ($l=200$nm, $a=20$nm) and a birefringent crystal separated by 266 nm. Considering noise from thermal interaction and photon recoil, we get the sensitivity of our system, which is on the order of $10^{-28}~\rm Nm/\sqrt{Hz}$ at $10^{-7}$ torr. Therefore, the system will allow us to measure the Casimir torque and test the fundamental prediction of quantum electrodynamics \cite{Parsegian, Barash, Enk1995, Casimir force measurement, Casimir force 1998, Casimir Decca 2003, repulsive Casimir force, Yukawa Decca 2016, Jacob book, casimir review, Munday, V.Mostepanenko}. Besides its fascinating origin, the QED torque between anisotropic surfaces is expected to be important for the anisotropic growth of some crystals \cite{Yasui2015} and biological membranes. Our system will enable many other precision measurements, such as a measurement of the torque on a single nuclear spin \cite{Li}. It can also study electrostatics of surfaces.

\vspace{0.3cm}
\section*{Acknowledgments}
 We thank C. Zhong, F. Robicheaux, E. Fischbach, Z.-Q. Yin, J. Ahn, and J. Bang  for helpful discussions. This work is partly supported by the National Science Foundation under grant No. PHY-1555035.

%\bibliography{apssamp}% Produces the bibliography via BibTeX.

\end{document}